\documentclass[aps, pre, twocolumn, a4]{revtex4-1}
\pdfoutput=1
\usepackage{amsmath}
\usepackage{amssymb}
\usepackage{graphicx}
\begin{document}

\title{Molecular simulation study of polar order in orthogonal bent core smectic liquid crystals  }
\author{Stavros D. Peroukidis}
\author{Alexandros G. Vanakaras}
\author{Demetri J. Photinos}
\affiliation{Department of Materials Science, University of Patras, Patras, 26504, Greece}
%\date{September 3, 2014}

\begin{abstract}
We explore the phase behavior and structure of orthogonal smectic liquid crystals consisting of bent-core molecules (BCMs) by means of Monte Carlo molecular simulations. A simple athermal molecular model is introduced that describes the basic features of the BCMs. Phase transitions between uniaxial and biaxial (antiferroelectric) orthogonal smectics are obtained. The results indicate the presence of local in-plane polar correlations in the uniaxial smectic phase. The macroscopic uniaxial-biaxial transformation is rationalized in terms of local polar correlations giving rise to polar domains. The size of these polar domains grows larger under the action of an external vector field and their internal ordering is enhanced, leading to field-induced biaxial order-disorder transitions.
\end{abstract}

\maketitle

%%%%%%%%%%%%%%%%%%%%%%%%%%%%%%%%%%%%%%%%%%%%%%%%%%%%%%%%
%%%%%%%%%%%%%%%%%%%%%%%%%%%%%%%%%%%%%%%%%%%%%%%%%%%%%%%%
%%%%%%%%%%% Here begins the letter %%%%%%%%%%%%%%%%%%%%%
%%%%%%%%%%%%%%%%%%%%%%%%%%%%%%%%%%%%%%%%%%%%%%%%%%%%%%%%
%%%%%%%%%%%%%%%%%%%%%%%%%%%%%%%%%%%%%%%%%%%%%%%%%%%%%%%%
\section{Introduction}

Bent-core  molecules (BCMs) are known \cite{Reddy2006,Takezoe2006,Hough2009,Jakli2013,*Eremin2014} to self organise into a variety of soft complex structures, often showing hierarchal ordering on different length scales. Of particular interest for electro-optical \cite{Niori1996} and non-linear optical devices \cite{Exteb2008} is the polar (ferroelectric or antiferroelectric) switching of nematic and smectic BCM liquid crystals (LCs). The tilted stacking of the BCMs in layers is rather common and is the defining characteristic of the tilted smectic (Sm\textit{C}) class of materials as well as of the so called tilted cybotactic nematic class (N$_{cybC}$), combining the presence local tilted molecular layering with macroscopic positional disorder \cite{Keith2010, *France2010, *France2014}. The tilted ordering, however, complicates the device architecture, due to the monoclinic (or lower) symmetry it confers to the materials. In addition, it gives rise to the formation of chevron structures and defects %\cite{Panarin2010}%
 which can be a serious disadvantage for the use of tilted LC materials in electro-optic devices. On the other hand, when the stacking of the molecules into layers is not tilted, defining the so called orthogonal LCs, whether of the smectic (Sm\textit{A}) class or the orthogonal cybotactic nematic (N$_{cybA}$), the aforementioned tilt-inflicted drawbacks are removed and such materials could be very advantageous if they show polar order and optical biaxiality \cite{Shaker2011}.

The conventional nomenclature for orthogonal smectic phases of BCM materials includes the Sm\textit{A}, the Sm\textit{AP$_{R}$}, the Sm\textit{AP$_{A}$}, the  Sm\textit{AP$_{RA}$} and the Sm\textit{AP$_{\alpha}$}. The Sm\textit{AP$_{A}$} was first observed by Eremin et. al. \cite{Eremin2001} and characterized as an orthogonal biaxial smectic phase exhibiting antiferroelectric switching behaviour and consisting of polar layers of antiferroelectric order. The  Sm\textit{AP$_{R}$} as well as the Sm\textit{AP$_{R\alpha}$} were studied by Pociecha et. al. \cite{Pociecha2003} who proposed that the polar directors of the layers are arranged randomly in the first case and are helicoidally modulated along the layer normal, in the second. A similar interpretation for the structure of  Sm\textit{AP$_{R}$} has been provided by Panarin et. al. \cite{Panarin2011}. A different model of the structrure of   Sm\textit{AP$_{R}$} has been proposed by Shimbo et. al. \cite{Shimbo2006} wherein the molecules are organized in polar clusters (domains) that are randomly distributed within the layers, giving rise to a macroscopically uniaxial and apolar state. Furthermore, the recently observed  Sm\textit{AP$_{RA}$} is suggested to consist of randomly aligned nonpolar domains of local antiferroelectric ordering \cite{Gomola2010}.

These different interpretations of the connection between local structure and the macroscopic behavior of BCM LCs have stimulated a lot of interest for theoretical research. Systematic attempts have been made to rationalize this behaviour using theory and computer simulations \cite{Sreen2013,Bates2006,Lansac2003,*Camp1999,Dewar2004,*Dewar2005,Memmer2002,*John1,*Johnson2,*Orlandi,Peroukidis2011,Pelaez2008,Osipov2014}, but there is still much to be done, especially on the local structure-macroscopic properties connection. In previous works, the BC molecules were modelled as an assembly of interacting sites in a bent configuration by joining i) two hard spherocylinders \cite{Lansac2003,*Camp1999}, ii) Lenard-Jones spheres \cite{Dewar2004,*Dewar2005}, iii) Gay-Berne particles \cite{Memmer2002,*John1,*Johnson2,*Orlandi}, and iv) soft spherocyliders  \cite{Peroukidis2011}. A fully atomistic simulation of BCM nematics has also been reported \cite{Pelaez2008}.

In the present work, we have introduced a hard-core model for the molecular interactions that permits us to model BCMs with relatively short arms and sharp bend angles. The phase behavior and local structure is rationalized in terms of the shape anisotropy together with athermal specific interactions \cite{Peroukidis2012}. Finally, the response to external stimuli is also discussed and compared with experimental findings.

\section{Molecular model and simulation details}

We have employed a coarse grain molecular model which captures two essential features of the BCMs; these are: (i) the anisotropic shape of the mesogenic bent core and (ii) the chemical differentiation between the aromatic bent core of the mesogen and its terminal alkyl chains. The bent core of the molecules (see Fig. 1) consists of two spherocylinders (c-segments), of aspect ratio $L^{*}=L/D$, jointed rigidly at their end caps to form the bend angle $\gamma$, where $D$ is the diameter of the cylinders and the hemispherical caps. Two terminal spherical segments (t-segments) of radius $R^{*}=R/D$ are attached tangentially at the free ends of the c-segments of the molecule. In this work we have focused on systems with $L^{*}=2$, $R^{*}=0.7$  and $\gamma=120^{o}$ in accordance with the size and shape of common bent core smectics. We have also investigated systems with different $L^{*}$, $R^{*}$  and $\gamma$   to examine the influence of these parameters on the phase behaviour. The chemical incompatibility of different molecular units has been modelled by introducing the following differentiations among the segmental interactions \cite{Peroukidis2012}: (i) the c-segments interact with all types of segments (c or t) via hard core repulsion on overlap and null interaction otherwise, and (ii) the interaction potential of a t-t pair of segments is null, irrespectively of mutual overlapping. Our results demonstrate directly that these minimal molecular features are sufficient to generate the experimentally observed polymorphism and phase sequences.

The phase behaviour and molecular organization of the so defined BCM systems is studied by means of Metropolis Monte Carlo simulations in the isobaric isothermal ensemble ($NpT$) using variable size simulation boxes with periodic boundary conditions \cite{AllenBook}. We have performed compression and expansion series  by systematically varying the pressure. Equilibration requires on the order of $2$x$10^6$ cycles and a further $5$x$10^5$-$1$x$10^6$ cycles to be used for the calculation of ensemble averages of quantities of interest. A MC cycle consists, on average, of \textit{N} trial attempts (translations, orientations and translations-orientations) of a randomly chosen particle and one volume change attempt. The temperature and pressure are expressed in reduced units. The reduced temperature is kept constant $T^{*}=k_BT/\epsilon=1$, where $T$ is the temperature, $k_B$ is the Boltzmann constant and $\epsilon$ is the unit of energy. The reduced pressure is given by $p^{*}=pD^{3}/k_BT$. The number density is defined as $\rho^{*}=ND^{3}/V$, where $N$ is the number of molecules and $V$ is the volume. The inter-molecular distances $r$ and correlation lengths $\xi$ are scaled by the diameter $D$, defining the reduced distances $r^{*}=r/D$, the correlation lengths $\xi^{*}=\xi/D$ and the respective reduced scattering vector magnitudes $q^{*}=2\pi/r^{*}$.

\section{Results and discussion}
\subsection{Phase behaviour and molecular organisation in the smectic phases}
Systems consisting of BCMs with $L^{*}=2$, $R^{*}=0.7$  and $\gamma=120^{o}$  (see Fig. 1), comprising from $N=845$ to $3364$ molecules have been simulated. To quantify the orientational order and to identify the principal axes frame of the simulated systems we have diagonalized the order-tensors \cite{Camp1999}b $\boldsymbol{Q}^{a}=\tfrac{1}{2N}\sum_{i=1}^{N}\left[3\left(\boldsymbol{\hat{a}}_{i}\cdot\boldsymbol{\hat{A}}\right)\left(\boldsymbol{\hat{a}}_{i}\cdot\boldsymbol{\hat{B}}\right)-\delta_{AB}\right]$, 
where $\boldsymbol{\hat{a}}$    $=(\boldsymbol{\hat{x}},\boldsymbol{\hat{y}},\boldsymbol{\hat{z}})$ represents any of the molecular axes and $\boldsymbol{\hat{A}},\boldsymbol{\hat{B}}=\boldsymbol{\hat{X}},\boldsymbol{\hat{Y}},\boldsymbol{\hat{Z}}$ denote the axes of the simulation box. The eigenvector associated with the largest positive eigenvalue of the three order-tensors is taken as the primary director $\boldsymbol{\hat{n}}$. 

At relatively low pressures, the order parameters $S^{a}=\left\langle \tfrac{1}{N}\sum_{i}P_{2}\left(\boldsymbol{\hat{a}}_{i}\cdot\boldsymbol{\hat{n}}\right)\right\rangle $ approximately vanish, which indicates the absence of orientational order. Here  $P_{2}\left(\boldsymbol{\hat{a}}_{i}\cdot\boldsymbol{\hat{n}}\right)$ denotes the 2nd Legendre polynomial of the direction of the $x$ or $y$ or $z$ molecular axis relative to the primary director. The biaxial order parameters are evaluated from the components of the tensor $D^{a,b}=\left\langle \tfrac{1}{2N}\sum_{i}\left[\left(\boldsymbol{\hat{a}}_{i}\cdot\boldsymbol{\hat{l}}\right)^{2}+\left(\boldsymbol{\hat{b}}_{i}\cdot\boldsymbol{\hat{m}}\right)^{2}-\left(\boldsymbol{\hat{a}}_{i}\cdot\boldsymbol{\hat{m}}\right)^{2}-\left(\boldsymbol{\hat{b}}_{i}\cdot\boldsymbol{\hat{l}}\right)^{2}\right]\right\rangle $, 
 $\boldsymbol{\hat{a}}_{i}\neq\boldsymbol{\hat{b}}_{i}$ are axes of the ith molecule  and $\boldsymbol{\hat{l}},\boldsymbol{\hat{m}}$ denote  the secondary principal axes of the phase.
 
 \begin{center}
   \begin{figure}[h!]
   \includegraphics[scale=1.1,natwidth=528,natheight=717]{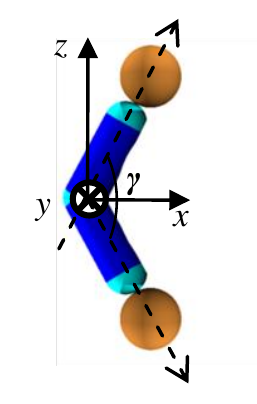}
   \caption{ Coarse-grain molecular model of BCM used in the simulations. The molecular axes frame is shown.}
   \label{fig:01}
   \end{figure}
 \end{center}

In Fig. 2(a) we present the calculated pressure vs number density  $p^{*}-\rho^{*} $ equation of state (EoS) and in Fig. 2(b) the the pressure dependence of the order parameters for a system of $N=845$ molecules. The system undergoes a transition, accompanied by a density jump, from the isotropic (\text{I}) to an orthogonal smectic (Sm\textit{A}) phase. A characteristic snapshot of this smectic phase is shown in Fig. 3a, from which it is evident that the molecular axis $z$ is ordered. The respective order parameter $S^{z}$ increases from nearly zero to approximately 0.9 (see Fig. 2(b)). At even higher pressures a transition to a biaxial smectic (Sm\textit{AP$_{A}$}) with antiferroelectric arrangement occurs. A characteristic snapshot is shown in Fig. 3(b). The biaxial order parameter of the Sm\textit{AP$_{A}$} is in the range $0.4<D^{x,y}<0.8$ (see Fig. 2(b)). The reverse process, i.e. expansion from the Sm\textit{AP$_{A}$} that has been obtained by compression, shows considerable hysteresis, with extended range of stability for the orthogonal smectic Sm\textit{A} phase, compared to the compression series. Finally, at lower pressures the Sm\textit{A} is destabilized in favour of the isotropic phase.

 \begin{center}
   \begin{figure}[h!]
   \includegraphics[scale=1.1,natwidth=560,natheight=717]{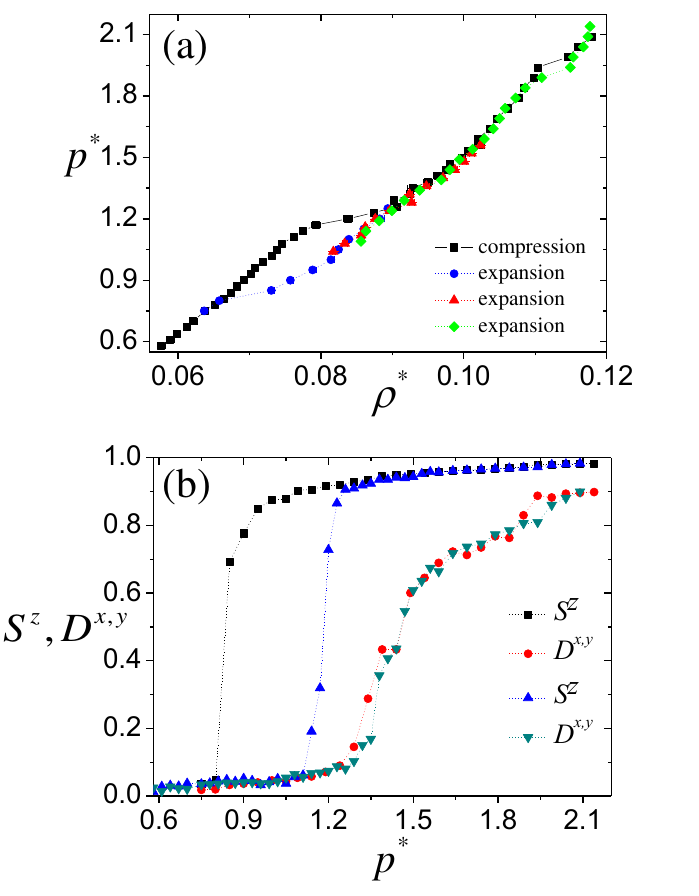}
   \caption{ (a) The equation of state for systems consisting of $N=$845 BCMs. (b) The order parameters $S^{z}$ and $D^{x,y}$ as a function of the reduced pressure $p^{*}$. Squares and circles correspond to expansion series from ordered phases and triangles to compression series from the isotropic phase. }
   \label{fig:02}
   \end{figure}
 \end{center}

 \begin{center}
   \begin{figure} [h!]
   \includegraphics[scale=1.0,natwidth=736,natheight=842]{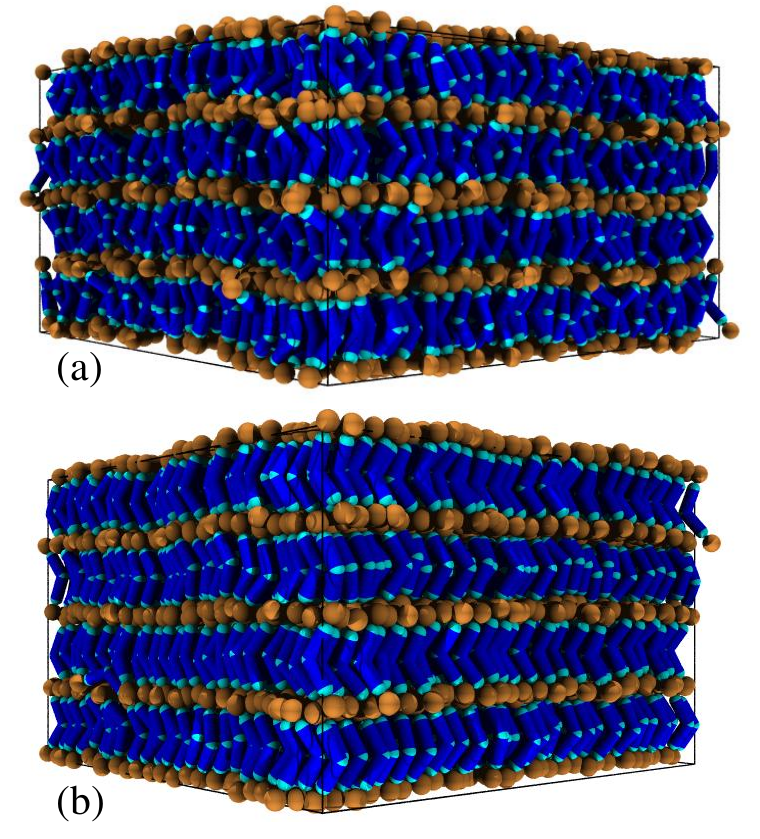}
   \caption{ Representative snapshots for the simulated BCM systems: (a) Sm\textit{A} at $p^{*}=1.1$ and (b) Sm\textit{AP$_{A}$} at $p^{*}=1.6$.}
   \label{fig:03}
   \end{figure}
 \end{center}

The structure of the isotropic liquid and of the ordered phases is examined through: (i) the calculated two dimensional x-ray scattering pattern \cite{Batesluck}; the intermolecular scattering intensity  $I_{inter}(\boldsymbol{q})$ reported in this study, is given by subtracting the single molecular intensity $I_{mol}(\boldsymbol{q})$  from the total intensity $I_{tot}(\boldsymbol{q})$, $I_{inter}(\boldsymbol{q})$= $I_{tot}(\boldsymbol{q})$ - $I_{mol}(\boldsymbol{q})$ with $I_{tot}(\boldsymbol{q})= f_{T}(\boldsymbol{q})f_{T}^{*}(\boldsymbol{q})$ and $I_{mol}(\boldsymbol{q})=\sum_{i=1}^{N}F^2_{j}\left(\boldsymbol{q}\right) $, where $f_{T}(\boldsymbol{q})$ is the total scattering (structure) factor $\sum_{i=1}^{N}F_{j}\left(\boldsymbol{q}\right)\exp\left(i\boldsymbol{q}\cdot\boldsymbol{r}_{j}\right)$,  $F_{j}\left(\boldsymbol{q}\right)$ is the spherical uniform scattering factor of the jth molecule, $\boldsymbol{r}_{j}$   is its position   and 
(ii) a set of positional two-dimensional pair correlation densities \cite{Peroukidis2011} defines as: 

$g_{0}^{\boldsymbol{\hat{a}},\boldsymbol{\hat{b}}}\left(r_{a},r_{b}\right)\thicksim \Big{\langle}\sum_{i\neq j}\delta\left(r_{a}-\boldsymbol{r}{}_{ij}\cdot\boldsymbol{\hat{a}}_{i}\right)\delta\left(r_{a}-\boldsymbol{r}{}_{ij}\cdot\boldsymbol{\hat{b}}_{i}\right)  $\\$\times\Theta\left[\left( \boldsymbol{r}{}_{ij}\cdot\left(\boldsymbol{\hat{a}}_{i}\times\boldsymbol{\hat{b}}_{i}\right)\right) ^2-\sigma^{2}\right] \Big{\rangle}_{i\neq j} $ $\\[3mm]$

which give the molecular density on the plane defined by the axes  of a single molecule. Here $\Theta(x)$ denotes the step-function ($\Theta=1$ for $x>0$  and $\Theta=0$  otherwise), and $\sigma=D/2$. 

First we study the structure of the thermodynamically stable isotropic phase (away from the hysteresis regime) which lacks long range positional and orientational order. The $g_{0}^{\hat{y},\hat{z}}\left(y,z\right)$ and $g_{0}^{\hat{x},\hat{z}}\left(x,z\right)$  show broad maxima (depicted by red arrows in Fig. 4) which are located at a distance slightly lower than one molecular length. The observation of these maxima indicates positional correlations along the $z$ molecular axis. A side by side packing of the molecular cores is evident from the intense maxima indicated by green arrows in Fig. 4(a).

It is interesting to note that, within the range of stability of the isotropic phase, the location of the maxima on the pattern does not change on increasing the pressure; only their intensity increases. The polar intermolecular correlations are rather weak and rapidly decay to zero with distance; this is evident from Fig. 4(c), showing the calculated orientational correlation function $g_{1}^{\hat{\boldsymbol{x}},\hat{\boldsymbol{y}}}\left(x,y\right)$, defined according to
$g_{1}^{\hat{\boldsymbol{a}},\hat{\boldsymbol{b}}}\left(r_{a},r_{b}\right)=\Big{\langle} \sum_{i\neq j}\left(\hat{\boldsymbol{x}}_{i}\cdot\hat{\boldsymbol{x}}_{j}\right)\delta\left(r_{a}-\boldsymbol{r}{}_{ij}\cdot\hat{\boldsymbol{a}}_{i}\right)\delta\left(r_{b}-\boldsymbol{r}{}_{ij}\cdot\hat{\boldsymbol{b}}_{i}\right) $\\$  \times\Theta\left[\left( \boldsymbol{r}{}_{ij}\cdot\left(\boldsymbol{\hat{a}}_{i}\times\boldsymbol{\hat{b}}_{i}\right)\right) ^2-\sigma^{2}\right]\Big{\rangle}_{i\neq j}/g_{0}^{\hat{\boldsymbol{a}},\hat{\boldsymbol{b}}}\left(r_{a},r_{b}\right).$ $\\[2mm]$

 Nevertheless, the nearest neighbours of a molecule in its $x-y$ plane appear to point in the same direction. The molecular arrangement described by the correlation functions in real space is also supported by the calculated x-ray scattering pattern. A spherical scatterer is assigned to the molecular apex (i.e. at the origin of the molecular axis frame). The inner ring (small angle region) corresponds to interlayer distance  $r^{*}=2\pi/q^{*}\thickapprox6.0$ in accordance with the intermolecular distance calculated from the density functions. Finally, the outer ring corresponds to the side-by-side molecular packing. In the isotropic phase the correlations are local, excluding any long range orientational or positional order. This, to our knowledge, is the first local structure analysis in the isotropic phase of BCMs systems using molecular simulations. Interestingly, the same smectic-like clusters, i.e domains of enhanced positional correlations, have been observed in real systems of bent core molecules in the isotropic phase \cite{Hong2010} and were invoked for the interpretation of interesting properties such as large flow birefringence \cite{Bailey2009}. At higher pressures the system undergoes a transition to an orthogonal smectic phase. This phase consists of apolar layers with spacing approximately equal to that observed in the local positional correlations.

 \begin{center}
   \begin{figure}[h!]
   \includegraphics[scale=1.0,natwidth=694,natheight=631]{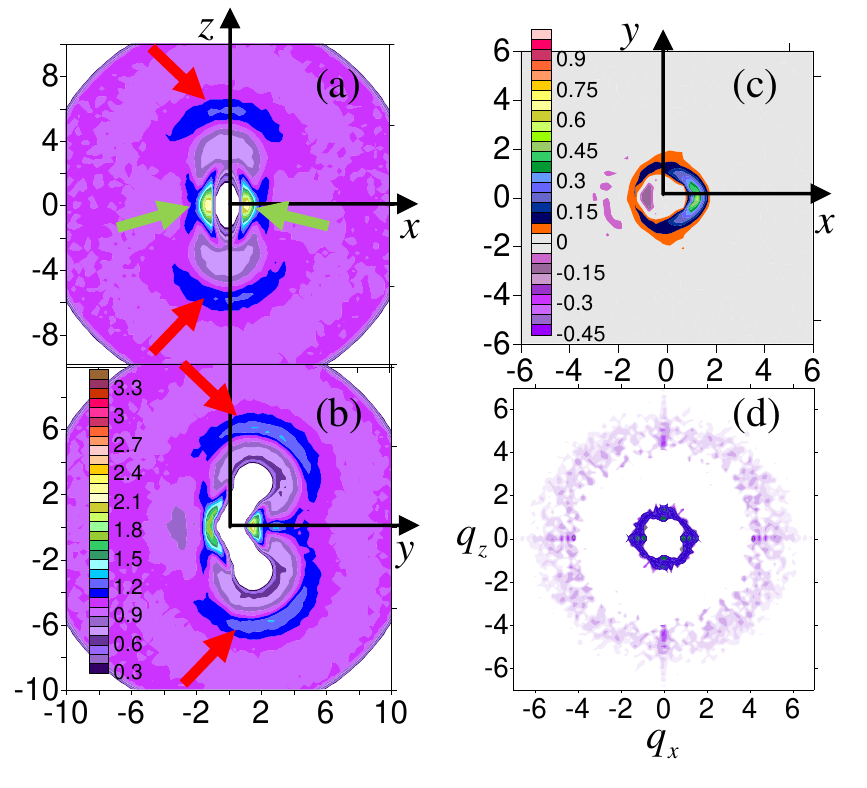}
   \caption{ Calculated correlation functions for systems of BCMs with $L^{*}=2$, $R^{*}=0.7$  and $\gamma=120^{o}$ in the isotropic phase at  $p^{*}=0.7$:   (a) $g_{0}^{\hat{x},\hat{z}}\left(x,z\right)$,  (b) $g_{0}^{\hat{y},\hat{z}}\left(y,z\right)$  in real space, (c) $g_{1}^{\hat{x},\hat{y}}\left(x,y\right)$  and (d) x-ray scattering pattern with scattering vector magnitude $q^{*}=2\pi/r^{*}$.}
   \label{fig:04}
   \end{figure}
 \end{center}
 
 In order to examine the in-plane polar order of the smectic phase we have evaluated the $g_{1}^{\hat{x},\hat{y}}\left(x,y\right)$ functions (see Fig. 5), from which we have found that short range polar correlations are present and that the range of orientational correlations is strongly anisotropic (extending in the $x$ direction more than in the $y$). Accordingly, each molecule in the layer can be viewed as being surrounded in the $x-y$  plane by an anisometric region containing neighbouring molecules that have biaxial correlations with that molecule. We have estimated the polar correlation lengths $\xi_{x},\xi_{y}$ along the $x$ and $y$ molecular axes from the $g_{1}^{\hat{x},\hat{y}}\left(x,y\right)$ functions as the optimal parameters for fitting the separate $x$ and $y$ dependences of this function to the functional forms $g_{1}^{\hat{x},\hat{y}}\left(x,0\right)=exp\left(-|x|/\xi_{x}\right)$ and $g_{1}^{\hat{x},\hat{y}}\left(0,y\right)=exp\left(-|y|/\xi_{y}\right)$. The correlation lengths  $\xi^*_{x}$ and $\xi^*_{y}$ increase with increasing pressure. These are roughly equal with very small values ($ \sim  1.0D$) just above the isotropic-smectic transition and increase to  $\xi^*_{x}=4.2$ and $\xi^*_{y}=2.5$   at pressures deep in the Sm\textit{A} phase (see Fig. 6). The calculated correlation lengths suggest that each polar domain incorporates up to a few tens of molecules.     
 
 Representative snapshots of the directions of the arrow vectors (identified with the unit vectors of the molecular $x$ axis in Fig. 1) within a smectic layer are shown in Figs. 7(a)-(b). There it can be seen that, in the Sm\textit{A} phase, small polar domains exist and are distributed randomly within the layer. Some polar domains are indicated in Fig. 7(a)-(b) using ellipsoids with major and minor radius of approximately the lengths $\xi^*_{x}$ and $\xi^*_{y}$, respectively.  The fluidity of particles and the isotropic positional distribution within the layers should be noted. Optical inspection of uncorrelated snapshots (every one thousand of MC cycles) show a continuous formation and disappearance of polar domains. This indicates that these domains can not be considered as stable floating "polar islands" within the layers. The stochastic nature of the Monte Carlo simulations do not allow inferences about the dynamics of the domain formation and disappearance process. The size of these domains increases with pressure (this is also reflected on the persistence of the $g_{1}^{\hat{x},\hat{y}}\left(x,y\right)$  function over longer intermolecular distances). Interestingly, the sign alternation of  $g_{1}^{\hat{x},\hat{z}}\left(x,z\right)$  with $z$ near the transition to the Sm\textit{AP$_{A}$} indicates that the small polar domains positioned one just above the other on successive layers show antiferroelectric order correlations (see Fig. 5(b)). Therefore, the Sm\textit{A} phase in this limit incorporates the characteristics of a conventional Sm\textit{A} phase and also consists of weakly correlated antiferroelectrically ordered domains. We term this smectic state as Sm\textit{A$_{Pc}$}. Note that, according to ref. \cite{Panarin2010}, the layers in a  Sm\textit{AP$_{R}$} are polar, with the polar directors randomly arranged across the different smectic layers, thus producing a macroscopically apolar smectic phase. This interpretation differs from the one given in ref. \cite{Shimbo2006,Gomola2010}, which is supported by our findings, i.e. each of the layers is overall apolar and consists of randomly oriented polar domains. Hence, the origin of the macroscopic apolarity of the phase is attributed to cancellations of local polar correlations (extending few molecular diameters) within each layer, as opposed to the mechanism of polarity cancellations among macroscopically polar layers.

 \begin{center}
   \begin{figure}[h!]
   \includegraphics[scale=0.85,natwidth=645,natheight=494]{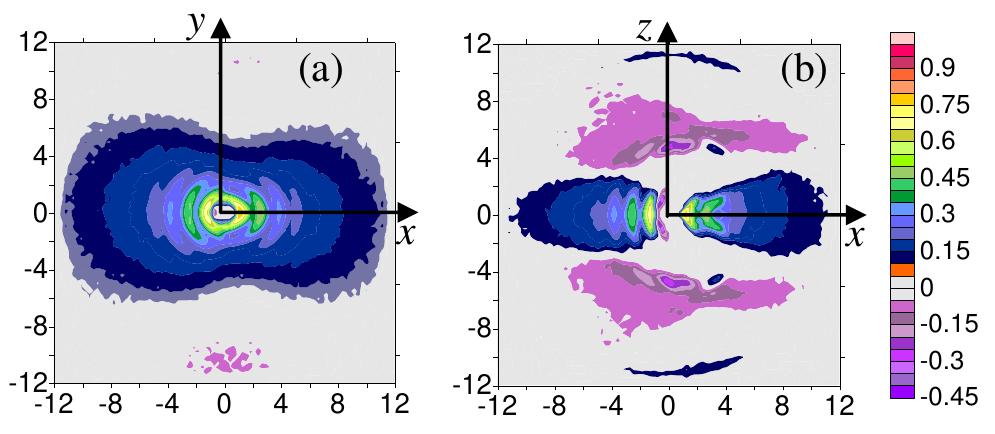}
   \caption{ Contour maps of calculated correlation functions for systems of BCMs with $L^{*}=2$, $R^{*}=0.7$  and $\gamma=120^{o}$,   (a) $g_{1}^{\hat{x},\hat{y}}\left(x,y\right)$,  (b) $g_{1}^{\hat{x},\hat{z}}\left(x,z\right)$  for the Sm\textit{A$_{Pc}$} phase at $p^{*}=1.30$.}
   \label{fig:05}
   \end{figure}
 \end{center}

 \begin{center}
   \begin{figure}[h!]
   \includegraphics[scale=0.80,natwidth=772,natheight=631]{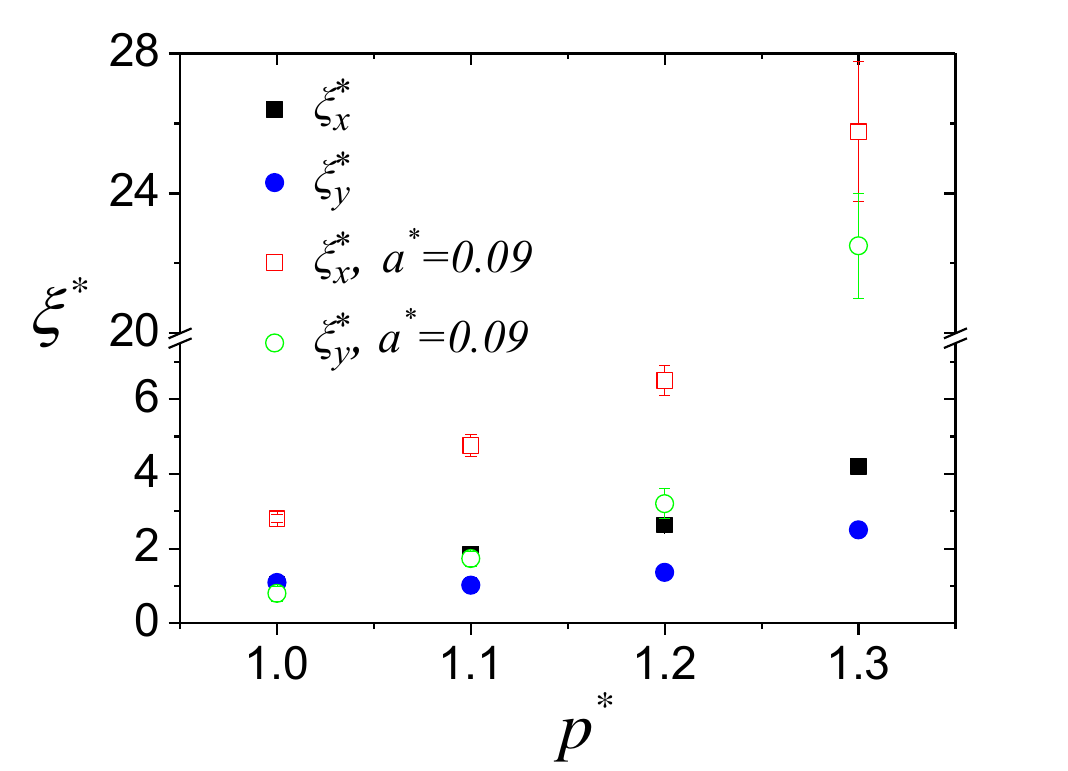}
   \caption{Calculated polar correlation lengths $\xi^*_{x}$ (squares) and $\xi^*_{y}$ (circles) vs pressure in the Sm\textit{A$_{Pc}$} phase. Solid symbols: without external field, open symbol: under external field $a^{*}=0.09$. }
   \label{fig:06}
   \end{figure}
 \end{center}
 
Further compression of the system brings it to a transition from the  Sm\textit{A$_{Pc}$} to the Sm\textit{AP$_{A}$} phase. In this phase, each layer becomes macroscopically polar, with  the direction of polarity  alternating from layer to layer, thus forming a macroscopic antiferroelectric structure.   This implies that the polar domains observed in Sm\textit{A$_{Pc}$} increase in size and merge upon the transition to the Sm\textit{AP$_{A}$} giving rise to long-range two-dimensional polar order within each layer. In this case the antiferroelectric coupling between adjacent smectic layers counterbalances the spontaneous tendency towards disorder in two-dimensional systems. We note here that the phase sequence: \textit{I}-Sm\textit{A}-Sm\textit{AP$_{R}$}-Sm\textit{AP$_{A}$}-\textit{Cr}  has recently been verified experimentally \cite{Panarin2010}; in our case, the  Sm\textit{A}-Sm\textit{AP$_{R}$} phase transition of \cite{Panarin2010}  appears as a nearly continuous transition from an orthogonal smectic with small polar domains to an orthogonal smectic with larger ones.

 \begin{center}
   \begin{figure}[h!]
   \includegraphics[scale=0.80,natwidth=556,natheight=549]{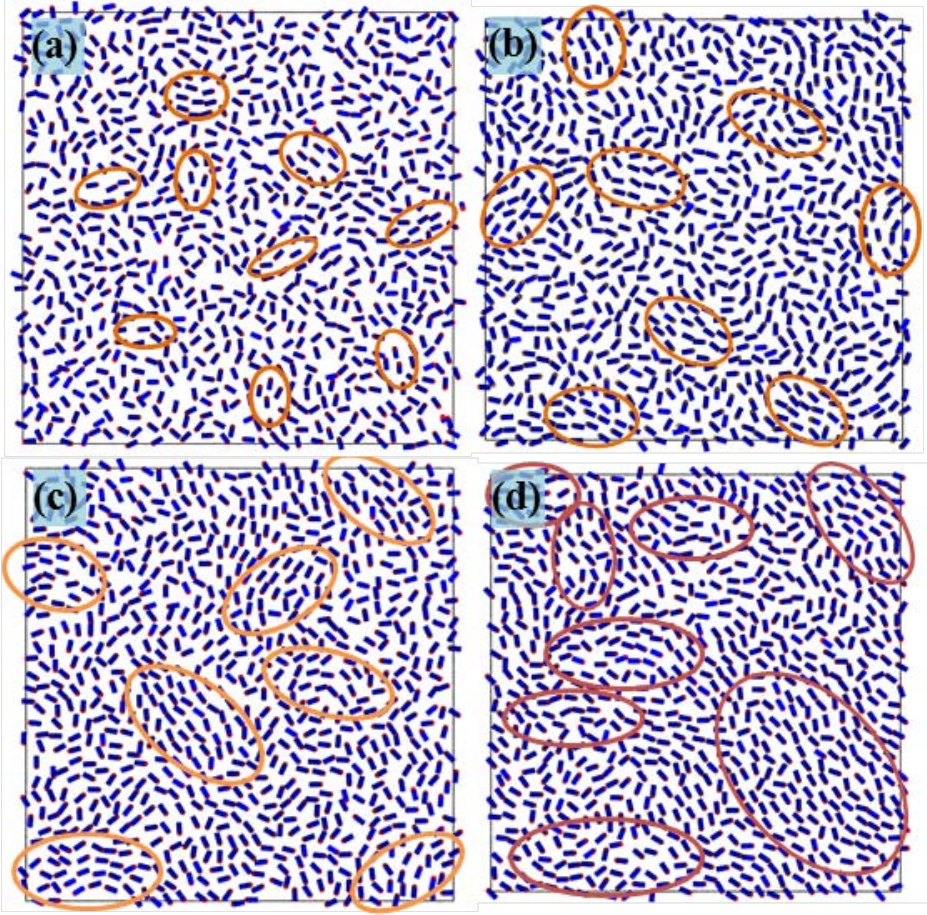}
   \caption{Representative snapshots presenting the molecular arrow vector (unit vector along the molecular $x$ axis) in a single smectic layer at two pressures for systems with (c,d) and without (a,b) external aligning field.   top panel, $a^*=0$: (a) Sm\textit{A$_{Pc}$}  phase at $p^{*}=1.20$ and (b) at $p^{*}=1.30$;  bottom panel: same pressures with $a^*=0.09$. The ellipses are drown to indicate some polar domains of the systems and to emphasise the larger polar correlation lengths in the field-on  systems.}
   \label{fig:07} 
   \end{figure}
  \end{center}

On increasing the bend angle to $\gamma=140^{o}$ or $160^{o}$  for the same $L^{*}=2$ and $R^{*}=0.7$, the range of stability of the  Sm\textit{A} phase increases. On the other hand, increasing the length of the arms to $L^{*}=3$ with $R^{*}=0.7$ and $\gamma=120^{o}$ destabilizes the Sm\textit{A} phase, which is now observed within a small range of pressures. Finally, a slight decrease of the size of the t-segments destabilizes the liquid crystalline phases: the system with $L^{*}=2$ and $R^{*}=0.5$  and $\gamma=120^{o}$ exhibits only an isotropic to crystal transition. 
\subsection{Domain structure in the presence of static alignment polar field}
A  macroscopic  measure of the size and internal ordering of the clusters can be provided by the susceptibility of the system to field-induced ordering \cite{Peroukidis2009,Vanakaras2008}. 
 A similar measure is provided by the threshold value of the applied field that is necessary to produce a transition from the state of randomly oriented clusters to the macroscopically ordered phase \cite{francescangeli_extraordinary_2011}. The transition mechanism in this case entails the merging and size increase of the clusters. For clusters identified on the basis of a vector property, such as the polar ordering of the molecules considered in these simulations, the appropriate susceptibility refers to the direct coupling of a vector property reflecting molecular polarity with a vector field. For electro-optic applications the response of the LC phase to the electric field is of primary interest and  therefore the relevant coupling would be that of a permanent molecular dipole moment with an applied electric field. However, as the aim of the present simulations is to identify the possible phase organization modes that originate directly from the bent shape of the molecules, the model molecules are not endowed with permanent electrostatic moments or polarisabilities. Strictly, therefore, the only dipole moments of the molecules are the “steric” ones, associated with their bent shape. In principle these would couple with mechanical vector fields producing, for example, polar flow-alignment. Alternatively, it may be formally assumed that the molecules carry permanent electric dipole moments that couple directly to an externally applied electric field but are very weak to generate significant intermolecular interactions. Generally, electric dipole interactions are known to influence the phase behavior and the local structure of LC systems~\cite{vanakaras_electric_1995, *berardi_dipole_2002}. Recently, a Sm\textit{A}-Sm\textit{AP$_{F}$} phase transition has been predicted theoretically \cite{Osipov2014} in a system of bent-core molecules possessing a permanent dipole moment along the arrow direction. An extension of the present molecular simulations is under way for the study of the phase behavior of systems of bent-core molecules possessing a permanent dipole, and of their response to an external electric field.  

In the present work we investigate the response of the system to an externally applied polar-field $\boldsymbol{\hat{E}}$ that couples linearly to the molecular arrow vector, $\boldsymbol{\hat{x}}$, and contributes to the potential energy of the ith molecule as $U_{i}=-a\left(\boldsymbol{\hat{x}}_{i}\cdot\boldsymbol{\hat{E}}\right)$, where $a$ is a coupling parameter. The dimensionless strength of the field is measured as $a^{*}=a/kT$.  When the field is off, the system does not show a net spontaneous polarity. This is clearly reflected through the vanishing first rank polar order parameter $P^{x}_{1}=\left\langle |\sum_{i=1}^{N}\boldsymbol{\hat{x}}_{i}\cdot\boldsymbol{\hat{m}}|/N\right\rangle$. Here $\boldsymbol{\hat{m}}$ denotes the secondary principal axis of the phase along which the molecular arrow-vectors order in the presence of the external field ($a^* \neq 0$); it coincides with the direction of the applied field.   
The dependence of biaxiality and polar order parameters on pressure, for various field strengths, are presented in Figs. 8(a) and 8(b) respectively. From these plots is easily concluded that fields with strengths $a^* \geq 0.06$ render the low-pressure Sm\textit{A$_{Pc}$} macroscopically polar and therefore biaxial. In contrast, a measurable macroscopic polarity in the Sm\textit{AP$_{A}$} phase is obtained only when the applied field exceeds a critical strength $a^* \approx 0.12$. Fields with strengths up to $a^*=0.12$ do not induce any detectable polarity in the crystalline phase ($p^*>1.9$). We stress here that the field strengths with $a^* \leq 0.12$ leave the calculated macroscopic biaxial order parameter of the Sm\textit{AP$_{A}$} phase practically unchanged with respect to the field-off calculations, see Fig. 8(a). In addition these fields do not affect the average density of the systems. These observations suggest that the chosen  strengths, $0.06 \leq
 a^* \leq 0.12$, correspond to relative weak fields since they do not influence the main order parameters of the unperturbed (field-off) systems.

 \begin{center}
   \begin{figure}[h!]
   \includegraphics[scale=1.3,natwidth=465,natheight=583]{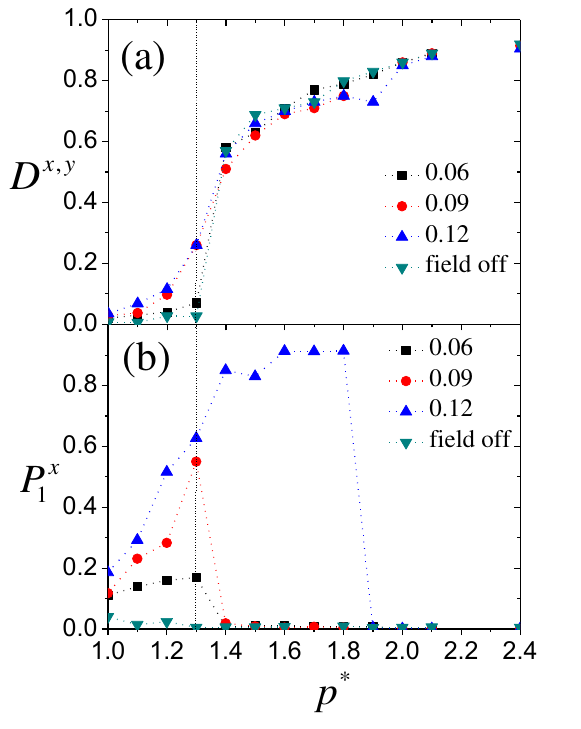}
   \caption{ (a) Biaxiality vs pressure and (b) Polar order parameter vs pressure  for various values of the field strength $a^{*}$ for a system of $N=3364$ molecules.  }
   \end{figure}
 \end{center}

Application of fields with strengths in the range $0.06 \leq a^{*} \leq 0.12$ to the Sm\textit{A$_{Pc}$} phase have the following consequences: i) the size of the polar domains, as quantified through the polar correlation lengths $\xi_x$ and $\xi_y$, increase with the strength of the field, see Figs. 6 and 7(c)-(d), ii) the development of orientational  correlations between the polar domains become long-ranged, leading to a net polarity and biaxiality and iii) the polar susceptibility of the system increases substantially as the system approaches the Sm\textit{A}$_{Pc}$-Sm\textit{AP}$_{A}$ phase transition at $p^{*}=1.30$.  

Up to this pressure --where the Sm\textit{A}$_{Pc}$-Sm\textit{AP}$_{A}$ phase transition occurs for the purely hardcore system without external field-- the macroscopic polarity of the filed-on system, as  reflected on $P^{x}_{1}$, grows continuously and reaches its maximum value. Above this pressure, depending on the strength of the applied field, we observe either an abrupt drop of the polarity (for $a^{*} \leq 0.09$) or, for higher fields, a continuous growth of $P^{x}_{1}$ with a tendency to saturate to its maximum allowed value as the systems reaches its crystallisation pressure. Clearly $a^{*} \approx 0.12$ is the lowest required field strength to have a field induced transition form an orthogonal  anti-ferroectically organised SmA phase (Sm\textit{AP}$_{A}$) into a ferroelectric orthogonal smectic (Sm\textit{AP}$_{F}$). Fields bellow this threshold leave the Sm\textit{AP}$_{A}$ phase practically unperturbed as indicated by the vanishing magnitude of the polar order parameter.   At even higher pressures ($p^{*}>1.90$) an antiferroelectric crystal is observed. In this case, much higher field strengths are expected for the formation of ferroelectric crystal state. 

Interestingly, the field response obtained in these simulation for the Sm\textit{AP} phase is in qualitative agreement to the experimentally observed \cite{Panarin2010} response to an external electric field for the Sm\textit{A}-Sm\textit{AP$_{R}$}  region. Our simulations suggest that the underlying microscopic picture of this transition involves polarly correlated in-plane domains that change their size and align under the influence of the external field, as opposed to randomly oriented macroscopically polar layers which are aligned by the external field. 

\section{Conclusions}

The thermodynamic stability and the local structure of liquid crystalline phases formed by model bent core molecules have been examined by the MC-$NpT$ simulation technique. The minimal representation of the intermolecular potential, through selective short-range athermal interactions, allows us to examine molecular structures with relative sharp angles and short arms without using the more complex parameterization of previous studies \cite{Dewar2004,*Dewar2005,Memmer2002,*John1,*Johnson2,*Orlandi,Peroukidis2011,Pelaez2008}.The molecular symmetry of the rigid bent core model we have used facilitates the investigation of  orthogonal smectic phases. The simple molecular model reproduces qualitatively experimental observations on orthogonal bent core systems \cite{Panarin2010}. We have observed the presence of local polar in plane domains which are randomly distributed in the Sm\textit{A$_{Pc}$} and their size increases on approaching the transition to the Sm\textit{AP$_{A}$}. Accordingly, the Sm\textit{A$_{Pc}$}-Sm\textit{AP$_{A}$} transition immerges from the growth of the size of these domains at the transition. The response of the uniaxial smectic phase to an external vector  field that couples to the polar ordering of the molecules is sensitively influenced by the presence of polar in-plane domains; these increase their size and get aligned by the field. Our results elucidate conflicting interpretations \cite{Panarin2010,Shimbo2006} regarding the microscopic mechanisms that are responsible for the macroscopic behavior of orthogonal bent core smectics especially for the  Sm\textit{A}-Sm\textit{AP$_{R}$}-Sm\textit{AP$_{A}$} transition.

\bibliography{MyReferences}

\end{document}